\documentclass[aps,nofootinbib,longbibliography,prd,twocolumn]{revtex4-1}
\usepackage[english,greek]{babel}
\usepackage[iso-8859-7]{inputenc}
\usepackage{graphicx}
\usepackage{teubner}
\usepackage{amsmath,amssymb}
\usepackage[colorlinks,citecolor=blue,linkcolor=blue,urlcolor=blue]{hyperref}
\usepackage{mathrsfs}
\usepackage{enumerate}
\def\be{\begin{equation}}
\def\ee{\end{equation}}

\usepackage[normalem]{ulem}

\begin{document}

\selectlanguage{english}

\title{Space and Time in Loop Quantum Gravity}
\author{Carlo Rovelli}
\affiliation{\small
\mbox{CPT, Aix-Marseille Universit\'e, Universit\'e de Toulon, CNRS, F-13288 Marseille, France.} }
\date{\small\today}

\begin{abstract}
\noindent 
Quantum gravity is expected to require modifications of the notions of space and time.  I discuss and clarify how this happens in Loop Quantum Gravity. \\
{\em [Written for the volume ``Beyond Spacetime: The Philosophical Foundations of Quantum Gravity" edited by 
Baptiste Le Biha, Keizo Matsubara and Christian Wuthrich.]}
\end{abstract}
\maketitle

\section*{I.\ \  Introduction}

Newton's success sharpened our understanding of the nature of space and time in the XVII century.  Einstein's special and general relativity improved this understanding in the XX century.  Quantum gravity is expected to take a step further, deepening our understanding of space and time, by grasping of the implications for space and time of the quantum nature of the physical world.

The best way to see what happens to space and time when their quantum traits cannot be disregarded is to look how this actually happens in a concrete theory of quantum gravity.  Loop Quantum Gravity (LQG) \cite{Rovelli:2004fk,Rovelli:2014ssa,ThiemannBook,Ashtekar:2012eb,Gambini,Gambini:2010kx,Perez:2012wv} is among the few current theories sufficiently developed to provide a complete and clear-cut answer to this question. 

Here I discuss the role(s) that space and time play in LQG and the version of these notions required to make sense of a quantum gravitational world.  For a detailed discussion, see the first part of the book \cite{Rovelli:2004fk}.  A brief summary of the structure of LQG is given in the Appendix, for the reader unfamiliar with this theory. 

\section*{II.\ \  Space}

Confusion about the nature of space --- even more so for time--- originates from failing to recognise that these are stratified,  multi-layered concepts.  They are charged with a multiplicity of attributes and there is no agreement on a terminology to designate spacial or temporal notions lacking same of these attributes.  When we say `space' or `time' we indicate different things in different contexts. 

The only route to clarify the role of space and time in quantum gravity is to ask what we mean \emph{in general} when we say `space' or `time' \cite{VanFraassen1985}.  There are distinct answers to this question; each defines a different notion of `space' or `time'. Let's disentangle them.  I start with space, and move to time, which is more complex, later on. 

\begin{description}

\item[\em Relational space] `Space' is the relation we use when we locate things. We talk about space when we ask ``\emph{Where} is Andorra?" and answer ``Between Spain and France".  Location is established in relation to something else (Andorra is located by Spain and France).  Used in this sense `space' is a relation between things. It does not require metric connotations. It is the notion of space Aristoteles refers to in his \emph{Physics}, Descartes founds on `contiguity', and so on.  In mathematics it is studied by topology.  This is a very general notion of space, equally present in ancient, Cartesian, Newtonian, and relativistic physics. 

{\em This notion of space is equally present in LQG}.   In LQG, in fact, we can say that something is in a certain location with respect to something else.   A particle can be at the same location as a certain quantum of gravity.  We can also say that two quanta are \emph{adjacent}.  The network of adjacency of the elementary quanta of the gravitational field is captured by the graph of a \emph{spin network} (see Appendix).  The links of the graph are the elementary adjacency relations. Spin networks describe relative \emph{spacial} arrangements of dynamical entities: the elementary quanta. 

\item[\em Newtonian space] In the XVII century, in the \emph{Principia}, Newton introduced a \emph{distinction} between two notions of space \cite{Newton1934}.  The first, which he called the ``common'' one, is the one illustrated in the previous item.   The second, which he called the ``true" one, is what has been later called Newtonian space.  Newtonian space is not a relation between objects: it is assumed by Newton to exist also in the absence of objects. It is an entity with no dynamics, with a metric structure: that of a 3d Euclidean manifold.   It is postulated by Newton on the basis of suggestions from ancient Democritean physics, and is essential for his theoretical construction.\footnote{During the XIX century, certain awkward aspects of this Newtonian hypostasis led to the development of the notion of `physical reference system': the idea that Newtonian space captures the properties of preferred systems of bodies not subjected to forces.  This is correct but already presupposes the essential ingredient: a fixed metric space, permitting to locate things  with respect to distant references bodies. Thus the notion of reference system does not add much to the novelty of the Newtonian ontology.}  Special relativity modifies this ontology only marginally, merging Newtonian space and time into Minkowski's spacetime.  

In quantum gravity, Minkowski spacetime and hence Newtonian space {\em appear only as an approximations}, as we shall see below. They have no role at all in the foundation of the theory.  

\item[\em General relativistic space] Our understanding of the actual physical nature of Newtonian space (and Minkowski spacetime)  underwent a radical sharpening with the  discovery of General Relativity (GR).  The empirical success of GR ---slowly cumulated for a century and recently booming--- adds much credibility to the effectiveness of this step.   What GR shows is that Newtonian space is indeed an entity as Newton postulated, but is not non-dynamical as Newton assumed.   It is a \emph{dynamical} entity, very much akin to the electromagnetic field: a gravitational field. Therefore in GR there are two distinct spacial notions.  The first is the simple fact that dynamical entities (all entities in the theory are dynamical) are localized with respect to one another (``This black hole is inside this globular cluster").    The second is a left-over habit from Newtonian logic: the habit of calling `space' (or `spacetime') one particular dynamical entity: the gravitational field.   There is nothing wrong in doing so,  provided that the substantial difference between these three notions of space (order of localization, Newtonian non-dynamical space, gravitational field) is clear.   

{\em LQG treats space (in this sense) precisely as GR does: a dynamical entity} that behaves as Newtonian space in a certain approximation.   However, in LQG this dynamical entity has the usual additional properties of \emph{quantum} entities. These are three: (i) Granularity.  The quantum electromagnetic field has granular properties: photons. For the same reason, the quantum gravitational field has granular properties: the elementary quanta represented by the nodes of a spin network. Photon states form a basis in the Hilbert state of quantum electromagnetism like spin network states form a basis in the Hilbert space of LQG.  (ii) Indeterminism. The dynamics of the `quanta of space' (like that of photons) is probabilistic.  (iii) Relationalism. Quantum gravity inherits all features of quantum mechanics including the weirdest. Quantum theory (in its most common interpretation) describes \emph{interactions} among systems where properties become actual.  So happens in LQG to the gravitational field: the theory describes how it interacts with other systems (and with itself) and how its properties become actual in interactions. More on this after we discuss time. 

\end{description}

\section*{III.\ \  Time}

The case with time is parallel to space, but with some additional levels of complexity \cite{Callender2011}. 

\begin{description}

\item[\em Relational time] `Time' is the relation we use when we locate events. We are talking about time when we ask ``\emph{When} shall we meet?" and  answer ``In three days".  Location of events is given with respect to something else. (We shall meet after three sunrises.)  Used in this sense time is a relation between events.   This is the notion Aristoteles refers to in his \emph{Physics}\footnote{The famous definition is: Time is \textgreek{\>{a}rijm\'os kin\'hsvews kat\`a t\`o pr\'oteron ka\`i \As{u}steron} ``The number of change with respect to before and after" (Physics, IV, 219 b 2; see also 232 b 22-23)\cite{Aristotle}.}, and so on.  It is a very general notion of time, equally present in ancient, Cartesian, Newtonian, and relativistic physics. 

{\em When used in this wide sense, `time' is definitely present in LQG}.   In LQG we can say that something happens \emph{when} something else happens.  For instance, a particle is emitted when two quanta of gravity join.  Also, we can say that two events are temporally adjacent.  A network of temporal adjacency of elementary processes of the gravitational field is captured by the \emph{spinfoams} (see Appendix).

\item[\em Newtonian time] In the \emph{Principia}, Newton distinguished two notions of time.  The first, which he called the ``common'' one, is the one in the previous item.   The second, which he called the ``true" one, is what has been later called Newtonian time.  Newtonian time is assumed to be ``flowing uniformly", even when nothing happens, with no influence from events, and to have a metric structure: we can say when two time intervals have equal duration.  Special relativity modifies the Newtonian ontology only marginally, merging Newtonian space and time into Minkowski spacetime.  

In LQG (Minkowsky spacetime and hence) Newtonian time {\em appears only as an approximation}. It has no role at all in the foundation of the theory.  

\item[\em General relativistic time] What GR has shown is that Newtonian time is indeed (part of) an entity as Newton postulated, but this entity is not non-dynamical as Newton assumed.   Rather, it is an aspect of a dynamical field, the gravitational field. What the reading $T$ of a common clock tracks, for instance, is a function of the gravitational field $g_{\mu\nu}$, 
\be
T=\int\sqrt{g_{\mu\nu}\; dx^\mu dx^\nu}.\label{propertime}
\ee 
In GR, therefore, there are two distinct kinds of temporal notions.  The first is the simple fact that all events are localized with respect to one another (``This gravity wave has been emitted \emph{when} the two neutron stars have merged", ``The binary pulsar emits seven hundred pulses \emph{during} an orbit").    The second is a left-over habit from Newtonian logic: the habit of calling `time' (in `spacetime') aspects of one specific dynamical entity: the gravitational field.  Again, there is nothing wrong in doing so, provided that the difference between  three notions of time (relative order of events, Newtonian non dynamical time, the gravitational field) are clear.   

{\em LQG treats time (in this sense) as GR does:} there is no preferred clock time, but many clock times measured by different clocks.  In addition, however, clock times undergo standard quantum fluctuations like any other dynamical variable.  There can be quantum superpositions between different values of the same clock time variable $T$.  

Our common intuition about time is profoundly marked by natural phenomena that are not \emph{generally} present in fundamental physics. Unless we disentangle these from the aspects of time described above, confusion reigns (I have extensively discussed the multiple aspects of temporality in the recent book \cite{Rovelli2018}).   These fall into two classes: 

\item[\em Irreversible time] When dealing with many degrees of freedom we recur to statistical and thermodynamical notions. In an environment with an  entropy gradient there are irreversible phenomena. The existence of traces of the past versus the absence of traces of the future, or the apparent asymmetry of causation and agency, are consequences of the entropy gradient (of what else?). Our common intuition about time is profoundly marked by these phenomena.  We do not know why was entropy as low in the past universe \cite{Earman2006}. (A possibility is that this is a perspectival effect due to the way the physical system to which we belong couples with the rest of the universe \cite{Rovelli2015}.) Whatever the origin of the entropic gradient, it is a fact that all irreversible phenomena of our experience can be traced to (some version) of it \cite{Reichenbach1958,Albert2000,Price}.  This has nothing to do with the role of time in classical or quantum mechanics, in relativistic physics or in quantum gravity.  There is no compelling reason to confuse these phenomena with issues of time in quantum gravity. 

Accordingly, nothing refers to `causation, `irreversibility' or similar, in LQG.   LQG describes physical happening, the way it happens, its probabilistic relations, the microphysics, not the statistics of many degrees of freedom, entropy gradients or related irreversible phenomena.  

To address these, and understand the source of the the  features that make a time variable `special', we need a general covariant quantum statistical mechanics.  Key steps in this direction exist (see \cite{Connes:1994hv,Rovelli:1993ys} on \emph{thermal time}, and \cite{Chirco2016} and references therein) but are incomplete.  They have no \emph{direct} bearing on LQG. 

\item[\em Experiential time]  The second class of  phenomena that profoundly affects our intuition of time  are those following from the fact that our brain is a machine that  (because of the entropy gradient) remembers the past and works constantly to anticipate the future \cite{Buonomano2017}.   This working of our brain gives us a  distinctive feeling about time:  this is the feeling we call ``flow", or the ``clearing" that is is our experiential time \cite{Heidegger1950}.  This depends on the working of our brain, not on fundamental physics \cite{James1890}.  It is a mistake to search something pertaining to our feelings uniquely in fundamental physics.  It would be like asking fundamental physics to directly justify the fact that a red frequency is more vivid to our eyes than a green one: a question asked the wrong chapter of science.  

Accordingly, nothing refers to ``flowing", ``passage" or the similar in LQG.   LQG describes physical happening \cite{Dorato2013a}, the way they happen, their probabilistic relations, not idiosyncrasies of our brain (or our culture \cite{Everett2008}).   

\end{description}

\section*{IV.\ \  Presentism or block universe?\\ A false alternative.}

An ongoing discussion on the nature of time is framed as an alternative between presentism and block universe (or eternalism).  This is a false alternative.  Let me get rid of this confusion before continuing. 

Presentism is the idea of identifying what is \emph{real} with what is present now, everywhere in the universe. Special relativity and GR make clear that an objective notion of `present' defined all over the universe is not in the physical world.  Hence there can be no objective universal distinction between past, present and future.   Presentism is seriously questioned by this discovery, because to hold it we have to base it on a notion of present that lacks observable ground, and this is unpalatable. A common response states that (i) we must therefore identify what is \emph{real} with the ensemble of all events of the universe, including past and future ones \cite{Putnam1967}, and (ii) this implies that, since future and past are equally real, the passage of time is illusory, and there is no becoming in nature \cite{McTaggart1908}.   

The argument is wrong. (i) is just a grammatical choice about how we decide to use the ambiguous adjective ``real", it has no content \cite{Austin1962,Quine1948}.   (ii) is mistaken because  it treats time too rigidly, failing to realise that time can behave differently from our experience, and still deserve to be called time.  

The absence of a preferred objective present does not imply that temporality and becoming are illusions. Events happen, and this we call `becoming', but their temporal relations form a structure richer than we previously thought.  We have to adapt our notion of becoming to this discovery, not discard it. 

There are temporal relations, but these are local and not global; more precisely, there is a temporal ordering but it is a partial ordering, and not a complete one. The universe is an ensemble of processes that happen, and these are not organised in a unique global order. In the classical theory,  they are organised in a nontrivial geometry.  In the quantum theory, in possibly more complex patterns. 

The expression ``real now here" can still be used to denote an ensemble of events that sit on the portion of a common simultaneity surface for a group of observers in slow relative motion; the region it pertains to must be small enough for the effects of the finite speed of light to be smaller than the available time resolution.  When these conditions are not met, the expression ``real now" simply makes no sense.  

Therefore the discovery of relativity does not imply that becoming or temporality are meaningless or illusory: it implies that they behave in a more subtle manner than in our pre-relativistic intuition. The best language for describing the universe remains a language of happening and becoming, not a language of being.  Even more so when we fold quantum theory in. 

\emph{LQG describes reality in terms of processes}.  The amplitudes of the theory determine probabilities for processes to happen.  This is a language of becoming, not being.   In a process, variables change value.  The quantum states of the theory code the possible set of values that are transformed into each other in processes.  

In simple words, the \emph{now}  is replaced by \emph{here and now}, not by a frozen eternity. 

Temporality in the sense of becoming is at the roots of the language of LQG.  But in LQG there is no preferred time variable, as I discuss in the next section. 

\section*{V.\ \  ``Absence of time" and relative evolution: time is not forzen}

What is missing in LQG is not becoming.  It is a (preferred) time variable.  

Let me start by reviewing the (different) roles of the coordinates in Newtonian physics and GR. Newtonian space is a 3d Euclidean space and Newtonian time is a uniform 1d metric line.   Euclidean space admits families of Cartesian coordinates $\vec X$ and the time line carries a natural (affine) metric coordinate $T$.   These quantities are tracked  by standard rods and clocks.  Rods and clocks are not strictly needed for localisation in time and space, because anything can be used for relative localisation, but they are convenient in the presence of a rigid background metric structure such as the Newtonian, or the special relativistic one.  

Rods and clocks are also useful in GR, but far less central. Einstein relayed on rods and clocks in the early days of the theory, but later realized that this was a mistake and repeatedly de-emphasized their role at the foundation of his theory. In fact, he cautioned against giving excessive weight to the fact that the gravitational field defines a geometry~\cite{Lehmkuhl2014a}.  He regarded this fact as a convenient mathematical feature and a useful tool to connect the theory to the geometry of newtonian space~\cite{Einstein_1921}, but the essential about GR is not that it describes gravitation as a manifestation of a Riemannian spacetime geometry; it is that it provides a field theoretical description of gravitation \cite{Einstein_Meaning}.  

GR's general coordinates $\vec x, t$ are devoid of metrical meaning, unrelated to rods and clocks, and arbiltrarilly assigned to events.  This is imposed by the fact that the dynamics of rods and clocks is determined by interaction with the gravitational field.  Therefore the general relativistic coordinates do not have the direct physical interpretation of Newtonian and special relativistic coordinates. To compare the theory to reality we have to find coordinate-invariant quantities. This generates some technical complication but is never particularly hard in realistic applications.   But the relativistic $t$ coordinates should not be confused with intuitive time, nor with clock time.  Clock time is computed in the theory by the proper time \eqref{propertime} along a worldline.  The reason is that this quantity counts, say, the oscillations of a mechanism following the worldline. Contrary to what often wrongly stated, this is not a postulate of the theory: it is a consequences of the equations of motion of the mechanism.   

Given two events in spacetime, the clock time separation between them depends on the worldline of the clock.  Therefore there is no single meaning  to the time separation between two events.   This does not make the notion of time inconsistent: it reveals it to be richer than our naive intuition.  It is a fact that two clocks separated and then taken back together in general do not indicate the same time.  Accord of clocks is an approximative phenomenon due to the peculiar environment in which we conduct our usual business. 

Due to the discrepancy between clocks, it makes no sense to interpret dynamics as evolution with respect to one particular clock, as Newton wanted.\footnote{Given two clocks that measure different time intervals between two events, it make no sense to ask which of the two is `true time': the theory simply allows us to compute the way each changes with respect to the other.} Accordingly, the dynamics of GR is not expressed in terms of evolution in a single clock time variable; it is expressed in terms of relative evolution between observable quantities (a detailed discussion is in Chapter 3 of \cite{Rovelli:2004fk}).   This fact makes it possible to get rid of the $t$ variable all-together, and express the dynamical evolution directly in terms of the relative evolution of dynamical variables (Chapter 3 of \cite{Rovelli:2004fk}).  Thus, special clocks or preferred spacial or temporal variables are \emph{not} needed in relativistic physics.

A formulation of \emph{classical} GR that does not employ the time variable $t$ at all is the Hamilton-Jacobi formulation \cite{Peres1962}. It is expressed uniquely in terms of the three metric $q_{ab}$ of a spacelike surfaces and defined by two equations
\be
D_a\frac{\delta S[q]}{q_{ab}}=0, \ \ \ \ G_{abcd}\frac{\delta S[q]}{q_{ab}}\frac{\delta S[q]}{q_{cd}}+\det q\, R[q]=0
\label{HJ}
\ee
where $G_{abcd}=q_{ac}q_{bd}+q_{ad}q_{bc}-q_{ab}q_{cd}$ and $R$ is the Ricci scalar of $q$.  Notice the absence of any temporal coordinate $t$. In principle, knowing the solutions of these equations is equivalent to solving the Einstein equations.  Here $S[q]$ is the Hamilton--Jacobi function of GR.  When $q$ is the 3-metric of the boundary of a compact region $R$ of an Einstein space, $S[q]$ can be taken to be the action of a solution of the field equations in this region.   It is the quantity connected to the LQG amplitudes as in \eqref{eq1}.

Absence of a time variable does not mean that ``time is frozen" or that the theory does not describe dynamics, as unfortunately is still heard. 

Equations   \eqref{HJ} provide indeed an equivalent formulation of standard GR and can describe the solar system dynamics, black holes, gravitational waves and any other  \emph{dynamical} process, where things become, without any need of an independent $t$ variable.  In these phenomena many physical variables change together, and no preferred clock or parameter is needed to track change. 

The same happens in LQG. The quantum versions of \eqref{HJ} formally determine the transition amplitudes between quantum states of the gravitational field. These can be coupled to matter and clocks.  Variables change together and no preferred clock variable is used in the theory.   

It is in this weak sense that it is sometimes said that ``time does not exist" at the fundamental level in quantum gravity.  This expression means that there is no time variable in the fundamental equations.  It does not mean that there is no change in nature.   The theory indeed is formulated in terms of probability amplitudes for \emph{processes}. 

\section*{VI.\ \  Quantum theory without Schr\"odinger equation}

Quantum mechanics requires some cosmetic adaptations in order to deal with the way general relativistic physics treats becoming. General relativistic physics describes becoming as evolution of variables that change together, any of them can be used to track change. No preferred time variable is singled out.   Quantum mechanics instead is commonly formulated in terms of a preferred independent clock variable $T$.  Evolution in $T$ is expressed either in the form of Schr\"odinger equation
\be
i\hbar\frac{\partial\psi}{\partial T}=H\psi
\ee
or as a dynamical equation for the variables
\be
\frac{dA}{d T}=i\hbar[A,H], 
\ee
where $H$ is the Hamiltonian operator. Neither of these equations is adapt to describe relativistic relative evolution.  The extension of quantum theory to the relativistic evolution is however not very hard, and has been developed by many authors, starting from Dirac. See for instance Chapter 5 of \cite{Rovelli:2004fk}, or \cite{Rovelli:2014ssa} or, on a slightly different perspective, the extensive work of Jim Hartle \cite{Hartle} on this topic.  

Like classical mechanics, quantum mechanics can be phrased as a theory of the probabilistic relations between the values of variables evolving together, rather than variables evolving with respect to a single time parameter.   The Schr\"odinger equation is then replaced by a Wheeler-DeWitt equation 
\be
C\psi=0,
\label{C}
\ee
or as a dynamical equation for the variables
\be
[A,C]=0
\ee
for a suitable Wheeler-DeWitt operator $C$.  Again: these equations do not mean that time is frozen or there is no dynamics.  They mean that the dynamics is expressed as joint evolution between variables, instead than evolution with respected to a single special variable.

Formally: the $\hbar\to 0$ limit of equation \eqref{C} is the second equation in \eqref{HJ}; given boundary values, \eqref{C} is formally solved by the  transition amplitudes $W$; these can be expressed as a path integral over fields in the region and in the $\hbar\to 0$ limit $W \sim e^{i\hbar \frac{S}{\hbar}}$, where $S$ is a solution to \eqref{HJ}.   These are formal manipulations. LQG provides a finite and well defined expressions for $W$, at any order in a truncation in the number of degrees of freedom. 

\section*{VII.\ \  Quantum process = Spacetime region}

Quantum theory does not describe how things `are'.   It describe quantum events that happen when systems interact \cite{Rovelli2017b}. We mentally separate a `quantum system', for a certain time interval, from the rest of the world, and describe the way this interact with its surroundings.  This peculiar conceptual structure at the foundations of quantum theory takes a surprising twist in quantum gravity.

In quantum gravity we identify the process of the `quantum system' with a \emph{finite spacetime region}.  This yields a remarkable dictionary between the relational structure of quantum theory and the relational structure of relativistic spacetime:\\

  \begin{tabular}{@{} ccc @{}}
\hline
    quantum transition & $\leftrightarrow$ & 4d spacetime region \\ 
        initial and final states & $\leftrightarrow$ & 3d boundaries \\ 
   interaction (`measurement') & $\leftrightarrow$ & continguity \\ 
\hline
  \end{tabular}\\[1mm]
   
\begin{figure}[h]
	\centering
{\includegraphics[height=3cm]{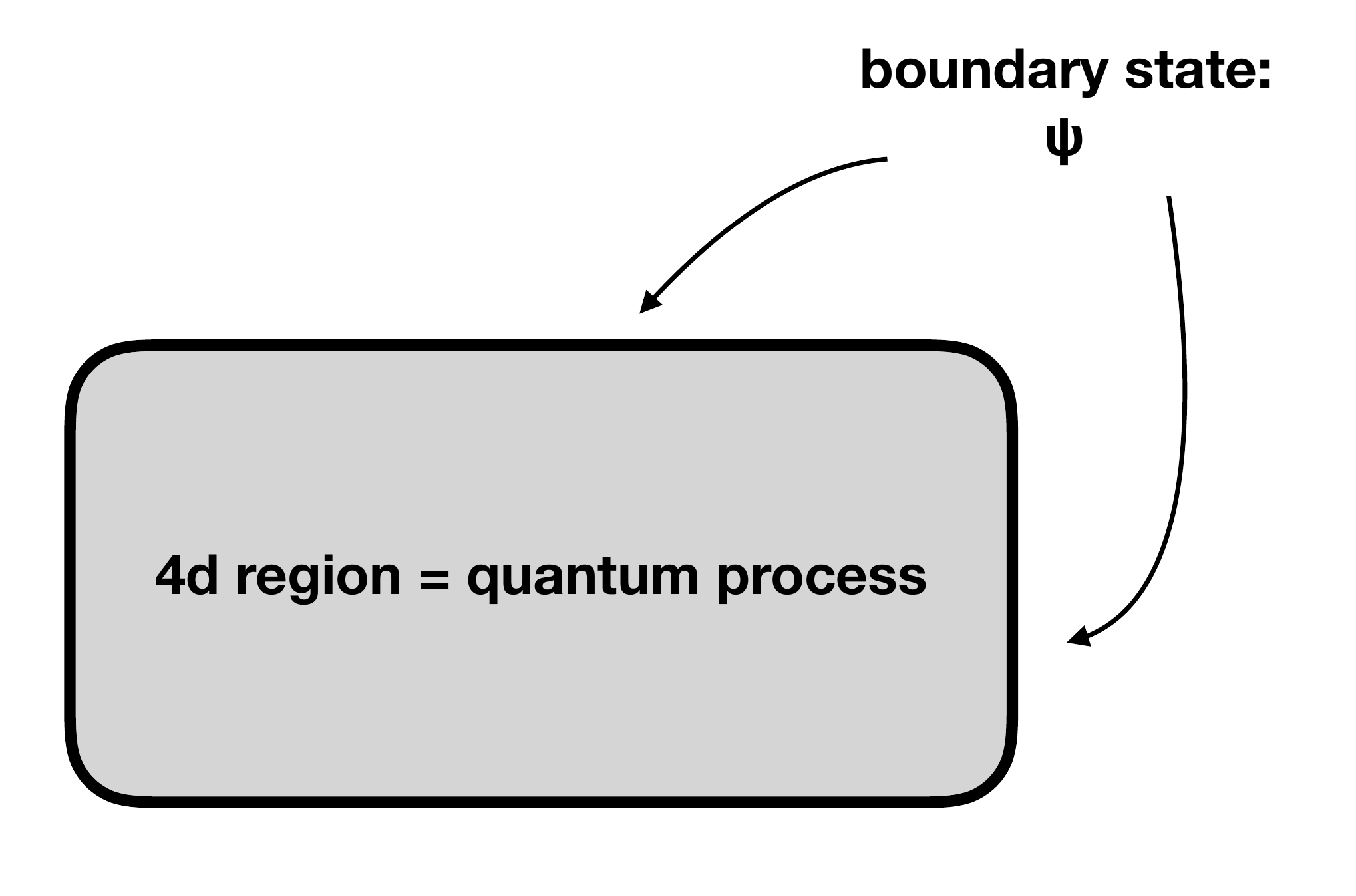}}
\caption{\em A compact spacetime region is identified with a quantum transition. The states of LQG sit on its boundary.}
\label{3}
\end{figure}

Thus the quantum states of LQG sit naturally on 3d boundaries of 4d regions (see Figure \ref{3}) \cite{Oeckl:2003vu}.  The quantum amplitudes are associated to what happens inside the regions.   Intuitively, they can be understood as path integral over all possible internal geometries, at fixed boundary data.    For each set of boundary data, the theory gives an amplitude, that determines the probability for this process to happen, with respect to other processes.

Remarkably: the net of quantum interactions between systems \emph{is the same thing} as the net of adjacent spacetime regions.

\section*{VIII.\  \ Conclusion}

`Space' and `Time' are expressions that can mean many different things:
\begin{enumerate}
\item Space can refer to the {\bf\em relative localisation} of things, time can refer to the {\bf\em becoming}  that shapes Nature.   As such, they are present in LQG like in any other physical theory.
\item Spacetime is a name given to the {\bf\em gravitational field} in classical GR.  In LQG there is a gravitational field, but it is not a continuous metric manifold.  It is a quantum field with  the usual quantum properties of discreteness, indeterminism and quantum relationality.
\item Space and time can refer to preferred variables used to locate things or to track change, in particular {\bf\em reading of meters and clocks}. In LQG, rods and clocks and their (quantum) behaviour can in principle be described, but play no role in the foundation of the theory.  The  equations of the theory do not have preferred spacial or temporal variables. 
\item Thermal, causal, {\bf\em ``flowing"} aspects of temporality are ground on chapters of science distinct from the elementary quantum mechanics of reality.  They may involve thermal time, perspectival phenomena, statistics, brain structures, or else. 
\item The universe described by quantum gravity is not flowing along a single time variable, nor organised into a smooth Einsteinian geometry.  It is a network of quantum processes,  related to one another, each of which obeys probabilistic laws that the theory captures. {\bf\em The net of quantum interactions between systems is identified with the net of adjacent spacetime regions}.
\end{enumerate}
These are the roles of space and time in Loop Quantum Gravity. Much confusion about these notions in quantum gravity is confusion between these different meanings of space and time. 

\appendix
\section{Loop Quantum Gravity in a nutshell}

\begin{figure}[h]
\includegraphics[height=2cm]{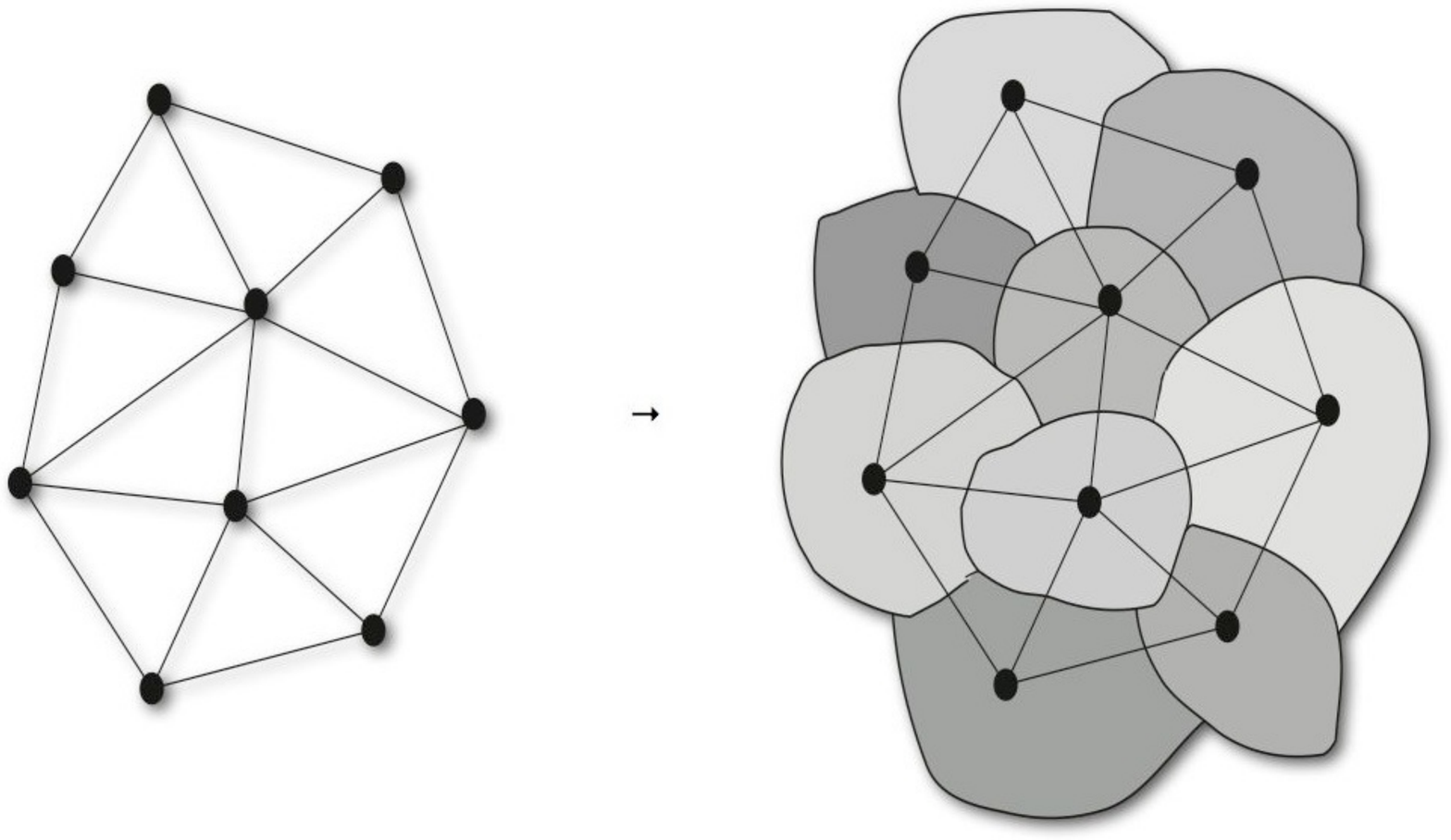}
\caption{\em The graph of a spin network and an intuitive image of the quanta of space it represents.}
\label{uno}
\end{figure}

As any quantum theory, LQG can be defined by a Hilbert space, an algebra of operators and a family of transition amplitudes. The Hilbert space $\cal H$ of the theory admits a basis called the \emph{spin network basis}, whose states $|\Gamma,j_l,v_n\rangle$ are labelled by a (abstract, combinatorial) {graph} $\Gamma$, a discrete quantum number $j_l$ for each link $l$ of the graph, and a discrete quantum number $v_n$ for each node $n$ of the graph. The nodes of the graph are interpreted as elementary `quanta of gravity' or `quanta of space', whose adjacency is determined by the links, see Figure \ref{uno}.  These quanta do not live on some space: rather, they themselves build up physical space. 

The volume of these quanta is discrete and determined by $v_n$. The area of the surfaces separating two nodes is also discrete, and determined by $j_l$. The elementary quanta of space do not have a sharp metrical geometry (volume and areas are not sufficient to determine geometry), but in the limit of large quantum numbers there are states in $\cal H$ that approximate 3d geometries arbitrarily well, in the same sense in which linear combinations of photon states approximate a classical electromagnetic field.  The spin network states are eigenstates of operators $A_l$ and $V_l$ in the operator algebra of the theory, respectively associated to nodes and links of the graph. In the classical limit these operators become functions of the Einstein's gravitational field $g_{\mu\nu}$, determined by the standard relativistic formulas for area and volume. For instance, $V(R)=\int_R \sqrt{\det q}$, for the volume of a 3d spacial region $R$, where  $q$ is the  3-metric induced on $R$. 

\begin{figure}[h]
\vspace{-2mm}
\includegraphics[height=2cm]{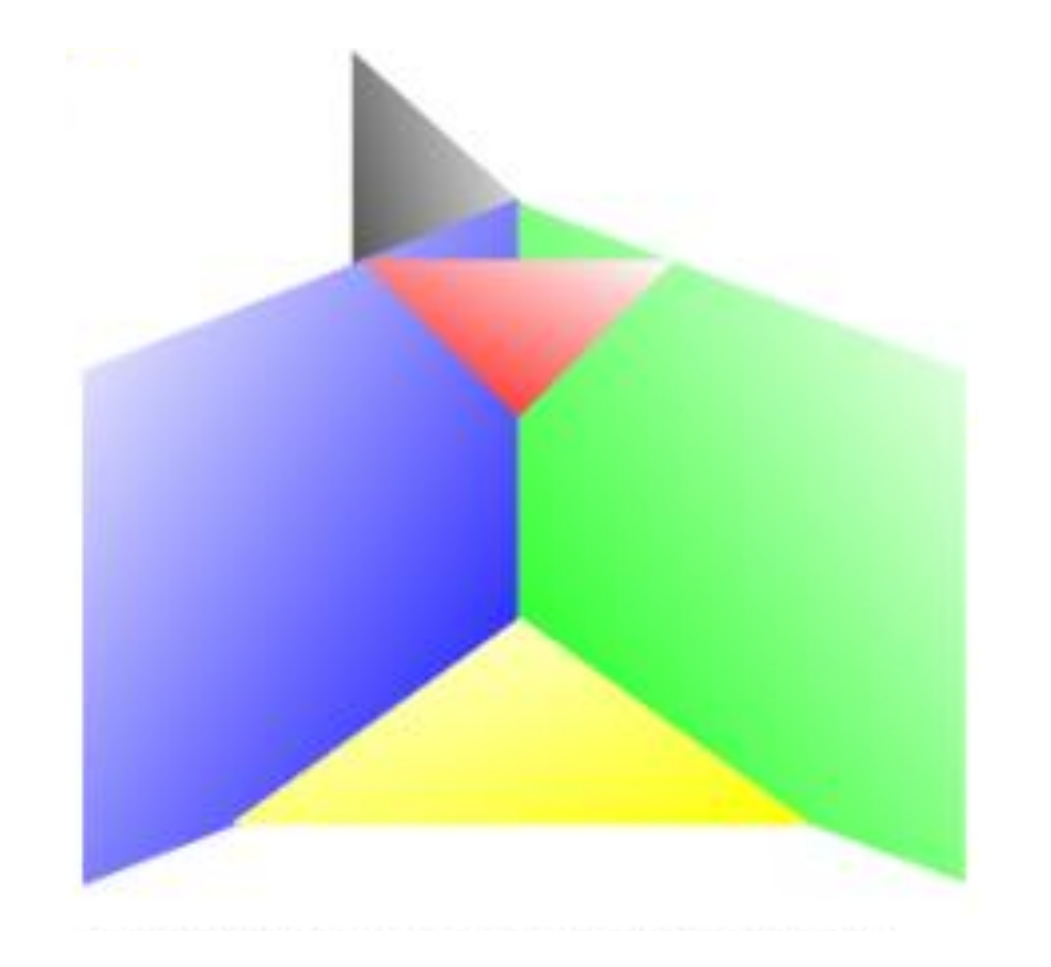}
\vspace{-3mm}
\caption{\em Spinfoam: the time evolution of a spin network.}
\label{due}
\end{figure}

In the covariant formalism (see \cite{Rovelli:2014ssa}), transition amplitudes are defined order by order in a truncation on the number of degrees of freedom.  At each order, a transition amplitude is determined by a \emph{spinfoam}: a combinatorial structure $\cal C$ defined by elementary faces joining on edges in turn joining on vertices (in turn, labeled by quantum numbers on faces and edges), as in Figure \ref{due}.

A spinfoam can be viewed as the Feynman graph of a history of a spin network; equivalently, as a (dual) discrete 4d geometry: a vertex corresponds to an elementary 4d region, an elementary process.  The boundary of a spinfoam is a spin network. The theory associates an amplitude $W_{\cal C}(\Gamma, j_l,v_n)$ (a complex number) to spinfoams. These are ultraviolet finite.  Several theorems relate them to the action (more precisely the Hamilton function $S$) of GR, in the limit of large quantum numbers.  This is the expected formal relation between the quantum dynamics, expressed in terms of transition amplitudes $W$ and its classical limit, expressed in terms of the action $S$: 
\be
W\sim e^{i\frac{S}{\hbar}}, 
\label{eq1}
\ee
where $W$ and $S$ are both functions of the boundary data. 

This concludes the sketch of the formal structure of (covariant) LQG.  Notice that nowhere in the basic equations of the theory a time coordinate $t$ or a space coordinate $x$ show up.

\providecommand{\href}[2]{#2}\begingroup\raggedright\endgroup


\end{document}